\documentclass[prl,showpacs,amsmath,amssymb,twocolumn]{revtex4}

\usepackage{graphicx}
\usepackage{dcolumn}
\usepackage{bm}
\usepackage{amssymb}   

\newcommand{\wen}{\ensuremath{W \rightarrow e \nu}}
\newcommand{\zee}{\ensuremath{Z \rightarrow ee}}
\newcommand{\wtaunu}{\ensuremath{W \rightarrow \tau \nu}}

\newcommand{\pte}{\ensuremath{p_T^e}}
\newcommand{\ptnu}{\ensuremath{p_T^\nu}}
\newcommand{\met}{\ensuremath{{\slash\kern-.7emE}_{T}}}
\newcommand{\vmet}{\ensuremath{\vec{\slash\kern-.7emE}_{T}}}

\begin{document}
\hspace{5.2in} \mbox{FERMILAB-PUB-09/449-E}
\title{Direct Measurement of the $W$ Boson Width}
%
\author{V.M.~Abazov$^{37}$}
\author{B.~Abbott$^{75}$}
\author{M.~Abolins$^{65}$}
\author{B.S.~Acharya$^{30}$}
\author{M.~Adams$^{51}$}
\author{T.~Adams$^{49}$}
\author{E.~Aguilo$^{6}$}
\author{M.~Ahsan$^{59}$}
\author{G.D.~Alexeev$^{37}$}
\author{G.~Alkhazov$^{41}$}
% ...

\affiliation{\vspace{0.1 in}(The D\O\ Collaboration)\vspace{0.1 in}}
% ...
\affiliation{$^{6}$University of Alberta, Edmonton, Alberta, Canada;
                Simon Fraser University, Burnaby, British Columbia, Canada;
                York University, Toronto, Ontario, Canada and
                McGill University, Montreal, Quebec, Canada}
% ...
\affiliation{$^{30}$Tata Institute of Fundamental Research, Mumbai, India}
% ...
\affiliation{$^{37}$Joint Institute for Nuclear Research, Dubna, Russia}
% ...
\affiliation{$^{41}$Petersburg Nuclear Physics Institute,
                St. Petersburg, Russia}
% ...
\affiliation{$^{49}$Florida State University, Tallahassee, Florida 32306, USA}
% ...
\affiliation{$^{51}$University of Illinois at Chicago,
                Chicago, Illinois 60607, USA}
% ...
\affiliation{$^{59}$Kansas State University, Manhattan, Kansas 66506, USA}
% ...
\affiliation{$^{65}$Michigan State University,
                East Lansing, Michigan 48824, USA}
% ...
\affiliation{$^{75}$University of Oklahoma, Norman, Oklahoma 73019, USA}
% ...
 
\date{September 25, 2009}

\begin{abstract}
We present a direct measurement of the width of the $W$ boson using the
shape of the transverse mass distribution of $\wen$ candidate events. 
Data from approximately 1 fb$^{-1}$ of integrated luminosity recorded
at $\sqrt{s}=1.96$ TeV by the D0 detector at the Fermilab Tevatron $p\bar{p}$
collider are analyzed. We use the same methods and data sample that 
were used for our recently published 
$W$ boson mass measurement, except for the modeling of the recoil, which is done with a 
new method based on a recoil library. Our result, $2.028 \pm 0.072$ GeV,
is in agreement with the predictions of the standard model.
\end{abstract}

\maketitle

The gauge structure of the standard model (SM) of electromagnetic, 
weak, and strong interactions tightly constrains 
the properties and interactions of the carriers
of these forces, the gauge bosons. Any departure from its
predictions would be an indication of physics beyond the SM.
The $W$ boson is one of the carriers of the weak force and has 
a predicted decay width of
\begin{equation}
\Gamma_W = (3+2 f_{QCD}) \frac{G_F M^3_W}{6\sqrt{2}\pi} (1+\delta),
\end{equation}
where $G_F$ is the Fermi coupling constant, $M_W$ is the mass of the 
$W$ boson and $f_{QCD}=3(1+\alpha_s(M^2_W)/\pi)$ 
is a QCD correction factor given to first order of the strong 
coupling constant $\alpha_s$.
The radiative correction $\delta$ is calculated to be 2.1\% 
with an uncertainty that is less than 0.5\% in the SM~\cite{theorywidth}.
Current world average values for $G_F$~\cite{pdg} and $M_W$~\cite{wmasscombo} 
predict $\Gamma_W = 2.093 \pm 0.002$ GeV. Physics 
beyond the SM, such as new heavy particles that couple to the $W$ boson,
could alter the higher order vertex corrections that enter into $\delta$ 
and modify $\Gamma_W$~\cite{{width_corr}}. 

Direct measurements of $\Gamma_W$ have been previously performed 
by the CDF and D0 collaborations~\cite{{CDFwidth},{CDFwidth2},{d0width},{combine}}. 
The width has also been directly measured at the CERN LEP $e^+e^-$ 
collider~\cite{{LEPwidth}}.
The combined Tevatron average is $\Gamma_W=2.056 \pm 0.062$ GeV, and the current 
world average is $\Gamma_W=2.106 \pm 0.050$ GeV~\cite{CDFwidth2}.

We present a direct measurement of $\Gamma_W$ using the 
shape of the transverse mass ($M_T$) distribution of $W\rightarrow e \nu$ 
candidates from  $p\bar{p}$ collisions with 
center-of-mass energy of 1.96 TeV using data from approximately 1 fb$^{-1}$ 
of integrated luminosity collected by the D0 detector~\cite{d0det}. The transverse mass is defined as
$M_T= \sqrt{2\pte \ptnu [1-\cos(\Delta \phi)]}$, where $\Delta \phi$ is 
the opening angle between the electron and neutrino in the plane perpendicular to the beam axis, and 
$p^e_T$ and $p^{\nu}_T$ are the transverse momenta of the electron and neutrino respectively. 
The fraction of events with large $M_T$ is sensitive to $\Gamma_W$,
although it is also influenced by the detector responses to the electron and the hadronic recoil.
We use a new data-driven method for modeling the hadronic recoil of the $W$ boson 
using a recoil library of $Z$ boson candidates~\cite{nim}. Aside from the recoil modeling, the method for extracting $\Gamma_W$ 
is similar to that described in a recent Letter on a measurement of $W$ boson mass by the D0 
collaboration~\cite{worldwmass}. 

The D0 detector includes a central tracking system, composed of a
silicon microstrip tracker (SMT) and a central fiber tracker,
both located within a 2~T superconducting solenoidal magnet and
optimized for tracking capability for $|\eta_D| \leq 3$~\cite{geo}. 
Three uranium and liquid argon calorimeters 
provide coverage for $|\eta_D| \leq 4.2$:
a central calorimeter (CC) covering $|\eta_D| \leq 1.1$, and two
endcap calorimeters (EC) with a coverage of $1.5 \leq |\eta_D| \leq 4.2$ 
for jets and $1.5 \leq |\eta_D| \leq 3.2$ for electrons.  
In addition to the preshower detectors, scintillators between the CC and EC 
cryostats provide sampling of developing showers at $1.1 \leq |\eta_D| \leq 1.5$.
A muon system surrounds the calorimetry and consists of three layers of scintillators
and drift tubes, and a 1.8~T iron toroid with a coverage of $|\eta_D| \leq 2$. 

The analysis uses $\wen$ candidates for the width extraction
and $\zee$ candidates to tune the simulation of the detector response used in the 
extraction of the $W$ boson width from data.  
The data sample was collected using a set of inclusive single-electron triggers.
The position of the reconstructed vertex of the hard collision 
along the beam line is required to be within 60 cm of the center of the detector.
Throughout this Letter we use ``electron" to imply either electron or positron.

Electron candidates are required to have $p^e_T>25$ GeV and must 
be spatially matched to a reconstructed track in the central 
tracking system. We calculate $p^e_T$ using the energy from the calorimeter and 
angles from the matched track.
The track must have at least one SMT hit and $p_T>10$ GeV. 
Electron candidates are further required to pass shower shape 
and energy isolation requirements and to be in the fiducial region
of the CC calorimeter. 

The neutrino transverse momentum, $\ptnu$, is inferred from the observed missing transverse energy, $\met$, 
reconstructed from $\vec{p}_T^{~e}$ and the transverse momentum of 
the hadronic recoil ($\vec{u}_T$) using $~\vmet = -[\vec{p}_T^{~e} + \vec{u}_T]$. 
The recoil vector $\vec{u}_T$ is the vector sum of energies in calorimeter 
cells outside those cells used for defining the electron. 
The recoil is a mixture of the ``hard" recoil that balances the boson transverse 
momentum and ``soft" contributions from particles produced by the spectator quarks, 
other $p\bar{p}$ collisions in the same beam crossing, electronics noise, and 
residual energy in the detector from previous beam crossings.

$W$ boson candidate events are required to have a CC electron with 
$\pte>25$ GeV, $~\met>25$ GeV, $u_T<15$ GeV, and $50<M_T<200$ GeV. 
$Z$ boson candidate events are required to have two CC electrons with $\pte>25$ 
GeV and $u_T<15$ GeV. These selections yield 499,830 $W$ boson candidates (5,272 candidates 
with $100<M_T<200$ GeV) and 18,725 $Z$ boson candidates with the 
invariant mass ($M_{ee}$) of the two electrons between 70 and 110 GeV.

The $W$ boson width is extracted by comparing the $M_T$ data distribution
with distributions in simulated templates generated at different width values.
The prediction (in number of events) of signal-plus-background is 
normalized to the data in the $50<M_T<100$ GeV window.
A binned negative log-likelihood method is used to
extract $\Gamma_W$ in the range $100<M_T<200$ GeV.

There are two main sources of events with high $M_T$: events
that truly contain a high mass $W$ boson, and events with a $W$ boson
whose mass is close to the $W$ boson mass central value but are 
produced with large $u_T$. This second category of events 
can be mis-reconstructed at high $M_T$ because
of resolution effects and also because the magnitude of the recoil
vector is systematically underestimated due to the response 
of the calorimeter to low energy hadrons, energy thresholds on the calorimeter 
energies, and magnetic field effects. 

Another experimental challenge arises from the $p_T$ dependence of the electron
identification efficiency, which can alter the shape
of the $M_T$ distribution. The electron isolation requirement used in this analysis
has a non-negligible dependence on the electron $p_T$ which is
measured using a detailed {\sc geant}-based Monte Carlo (MC) simulation~\cite{bib:geant} 
and tested using $\zee$ events. 

A fast MC simulation is used for the production of the $M_T$ templates. 
$W$ and $Z$ boson production and decay properties are modeled by 
the {\sc resbos} event generator~\cite{resbos} interfaced with {\sc photos}~\cite{photos}. 
{\sc resbos} uses gluon resummation at low boson $p_T$ and a next-to-leading order perturbative 
QCD calculation at high boson $p_T$. 
The CTEQ6.1M parton distribution functions (PDFs)~\cite{{cteq}} are used. {\sc photos} 
is used for simulation of final state radiation (FSR). 
Photons and electrons that are nearly collinear are merged
using an algorithm that mimics the calorimeter clustering algorithm. 

The detector response for electrons and photons, including energy calibration, showering 
and energy loss models, is simulated using a parameterization
based on collider data control samples, a detailed {\sc geant}-based 
simulation of the detector, 
and external constraints, such as the precise measurement of the $Z$ boson 
mass from the LEP experiments~\cite{lepzmass}. The primary control sample
is $\zee$ events, although $W \rightarrow e\nu$ events are also used in a limited way.
The modeling of the electron energy response, resolution and 
selection efficiencies is described in~\cite{worldwmass}.
The number of $Z$ boson candidates in data sets the scale for the systematic
uncertainties related to the electron modeling in the simulation, which are listed in detail in Table~\ref{t:syst}.

The modeling of the recoil is based on the recoil library obtained from $\zee$ events~\cite{nim}. 
A Bayesian unsmearing procedure~\cite{ago} allows the transformation 
of the two-dimensional distribution of reconstructed $Z$ boson $\vec{p}_T$ 
and the measured recoil momentum $\vec{u}_T$ to 
one between the true $Z$ boson $\vec{p}_T$ and the measured recoil $\vec{u}_T$. 
For each simulated $\wen$ event with a generator-level transverse momentum value $\vec{p}_T$, 
we select $\vec{u}_T$ randomly from the $Z$ boson recoil library with the 
same value of $\vec{p}_T$. The uncertainty on the recoil system simulation from 
this method is dominated by the limited statistics of the $Z$ boson sample; 
other systematic uncertainties originate from the modeling of FSR photons, acceptance differences 
between $W$ and $Z$ boson events, corrections for underlying energy beneath the electron cluster, 
residual efficiency-related correlations between
the electron and the recoil system, and the unfolding procedure.
Previous $M_W$ and $\Gamma_W$ measurements have relied upon 
parameterizations of the recoil kinematics based on phenomenological 
models of the recoil and detector response. The 
library method used here includes the actual detector response for
the hadronic recoil and also the correlations between different 
components of the hadronic recoil. 
This method does not rely on the {\sc geant}-based simulation of the
recoil system and does not have any tunable parameters.
The overall systematic uncertainty on $\Gamma_W$ due to the recoil 
model is found to be 41 MeV~\cite{nim}.

The backgrounds to $\wen$ events are (a) $\zee$ events in which one 
electron is not detected; (b) multijet production in which one jet is misidentified
as an electron and mis-measurement of the hadronic activity in the event
leads to apparent $\met$; (c) $\wtaunu \rightarrow e\nu\nu\nu$ events.
The $\zee$ background arises mainly when one of the two electrons is in the region between 
the CC and EC calorimeters. It is estimated from events with one electron with a high-$p_T$ 
track opposite in azimuth pointing towards the gap. 
The estimated background fraction is ($0.90 \pm 0.01$)\% for $50<M_T<200$ GeV.
The background fraction from multijet events is estimated from a
loose sample of candidate events without track match requirements and
then selecting a subset of events which satisfy the final tighter 
track match requirement. From $\zee$ events, and a sample of multijet
events passing the preselection but with low $\met$, we determine the
probabilities with which real and misidentified electrons will pass the 
track match requirement. These two probabilities, along with the
numbers of events selected in the loose and tight samples allow us to
calculate the fraction of multijet events in the dataset~\cite{matrix_method}.
The background contamination from multijet events is estimated
to be ($1.49 \pm 0.03$)\% for $50<M_T<200$ GeV. The 
$\wtaunu \rightarrow e\nu\nu\nu$ background is determined using a 
{\sc geant}-based simulation to be ($1.60 \pm 0.02$)\% for $50<M_T<200$ GeV 
and is normalized to the $\wen$ events in the same simulation.  
The overall background fraction is found to be $(4.36 \pm 0.05)\%$ 
with $M_T$ between 100 and 200 GeV. The uncertainties on the 
normalization and shape of the backgrounds cause a 6 MeV systematic 
uncertainty on $\Gamma_W$.

The systematic uncertainties in the determination of the $W$ boson width 
are due to effects that could alter the $M_T$ distribution.  
Uncertainties in the parameters of the fast MC simulation can affect 
the measurement of $\Gamma_W$. To estimate the effects, we allow 
these parameters to vary by one standard deviation and regenerate the $M_T$ 
templates. Systematic uncertainties 
resulting from the boson $p_T$ spectrum are evaluated by varying the $g_2$ parameter 
of the {\sc resbos} nonperturbative prescription within the uncertainties 
obtained from a global fit~\cite{g2} and propagating them to the $W$ boson width. 
Systematic uncertainties due to the PDFs are evaluated using the prescription 
given by the CTEQ collaboration~\cite{cteq}. Systematic uncertainties 
from the modeling of electroweak radiative corrections are obtained by 
comparisons with {\sc wgrad}~\cite{wgrad} and {\sc zgrad2}~\cite{zgrad}. 
The systematic uncertainty due to the $M_W$ uncertainty is obtained
by varying the input $M_W$ by $\pm 23$ MeV~\cite{wmasscombo}.

\begin{figure*}[tbp]
\includegraphics[scale=0.6]{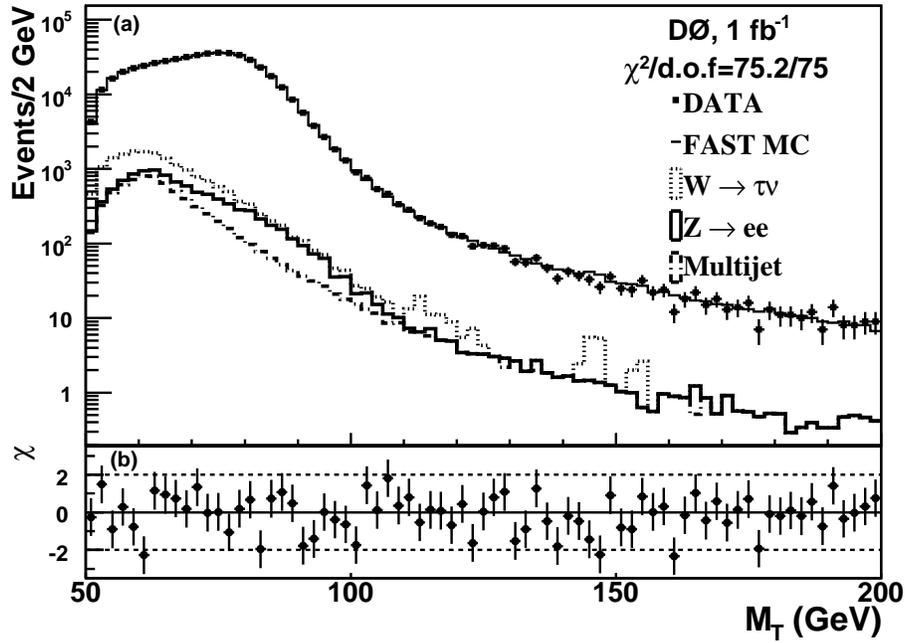}
\caption{\label{fig:finalresult}
Comparison of the $M_T$ data distribution with its expectation from a fast MC 
simulation of $W \rightarrow e\nu$ events to which smaller backgrounds have been added (a); 
$\chi$ values for each $M_T$ bin (b). The measured 
$\Gamma_W$ value is used for the fast MC prediction. 
The distribution of the fast MC simulation, including the cumulative contributions of the different backgrounds, is normalized to the
data in the region $50<M_T<100$ GeV.
}
\end{figure*}

We fit the $M_T$ data distribution to a set of templates generated with 
an input $W$ boson mass of 80.419 GeV at different assumed widths between a lower $M_T$ value and $M_T=200$ GeV. 
The lower $M_T$ cut is varied from 90 to 110 GeV to demonstrate 
the stability of the fitted result. 
While the statistical uncertainty decreases as the lower $M_T$ cut is 
reduced, the systematic uncertainty increases. The lowest overall
uncertainty is obtained for a lower $M_T$ cut of 100 GeV yielding 
$\Gamma_W = 2.028 \pm 0.039~(\mbox{stat}) \pm 0.061~(\mbox{syst})$ GeV. 
The $M_T$ distributions for the data and the MC template with backgrounds
for the best fit value are shown in Fig.~\ref{fig:finalresult}, which 
also shows the bin-by-bin $\chi$ values defined as the 
difference between the data and the template divided by the data statistical uncertainty. 

The methodology used to extract the width in this Letter is tested 
using $W$ and $Z$ boson events produced by a {\sc pythia}/{\sc geant}-based 
simulation and the same analysis methods used for the data.
The fast MC simulation is separately tuned for this study. 
Good agreement is found between the fitted $\Gamma_W$ value and the 
input $\Gamma_W$ value within the statistical precision of the test. 

The $\Gamma_W$ result obtained using the $M_T$ spectrum is 
in agreement with the predictions of the SM. 
We get consistent values of the $W$ boson width from fits to the $\pte$ 
distribution ($2.012 \pm 0.046~\mbox{(stat)}$ GeV) and the $\met$ 
distribution ($2.058 \pm 0.036~\mbox{(stat)}$ GeV). 
The width can also be estimated directly from the fraction
of events with $M_T>$ 100 GeV, and this gives $\Gamma_W=2.020 \pm 0.040~\mbox{(stat)}$ GeV.
The results are stable within errors when the data sample is divided into
different regions of instantaneous Tevatron luminosity, run epoch, and different
restrictions on $u_T$, electron $\eta_D$, $\vec{u}_T \cdot \hat{p}_T(e)$ and fiducial cuts on electron azimuthal angle.

\begin{table}
\begin{center}
  \begin{tabular}{lc} \hline \hline
   Source & $\Delta \Gamma_W$ (MeV) \\
  \hline
  Electron response model           & 33 \\
  Electron resolution model         & 10 \\
  Hadronic recoil model             &  41  \\
  Electron efficiencies             &  19  \\
  Backgrounds                       &  6 \\ 
  PDF                               &  20 \\
  Electroweak radiative corrections          &  7 \\
  Boson $p_T$                       &  1 \\ 
  $M_W$                             &  5 \\ 
  Total Systematic                  &  61 \\ \hline \hline 
  \end{tabular}
    \caption{Systematic uncertainties on the measurement of $\Gamma_W$.  
    \label{t:syst}}
\end{center}
\end{table}    

As a further cross check of the recoil library method we also use it to measure the $W$ boson mass using the
$M_T$ distribution over the region $65<M_T<90$ GeV. A value of $M_W = 80.404 \pm 0.023~(\mbox{stat})
\pm 0.038~(\mbox{syst})$ GeV is found, in good agreement 
with the result, $M_W = 80.401 \pm 0.023~(\mbox{stat})
\pm 0.037~(\mbox{syst})$ GeV, obtained using the same data set 
and the parameterized recoil model~\cite{worldwmass}.  

In conclusion, we have presented a new direct measurement of the width of the $W$ boson  
using 1 fb$^{-1}$ of data collected by the D0 detector at the Tevatron collider. 
A method to simulate the recoil system in $\wen$ events using 
a recoil library built from $\zee$ events is used for the first time. Our result, 
$\Gamma_W=2.028 \pm 0.039~\mbox{(stat)} \pm 0.061~\mbox{(syst)} = 2.028 \pm 0.072$ GeV, is in agreement with the prediction of the SM 
and is the most precise direct measurement result from a single experiment to date.

%
We thank the staffs at Fermilab and collaborating institutions,
and acknowledge support from the
DOE and NSF (USA);
CEA and CNRS/IN2P3 (France);
FASI, Rosatom and RFBR (Russia);
CNPq, FAPERJ, FAPESP and FUNDUNESP (Brazil);
DAE and DST (India);
Colciencias (Colombia);
CONACyT (Mexico);
KRF and KOSEF (Korea);
CONICET and UBACyT (Argentina);
FOM (The Netherlands);
STFC and the Royal Society (United Kingdom);
MSMT and GACR (Czech Republic);
CRC Program, CFI, NSERC and WestGrid Project (Canada);
BMBF and DFG (Germany);
SFI (Ireland);
The Swedish Research Council (Sweden);
Graduate Research Board, University of Maryland (USA);
and
CAS and CNSF (China).


\begin{thebibliography}{99}
%
\bibitem[a]{alton}
Visitor from Augustana College, Sioux Falls, SD, USA.
\bibitem[b]{atramentov,gershtein}
Visitor from Rutgers University, Piscataway, NJ, USA.
\bibitem[c]{burdin}
Visitor from The University of Liverpool, Liverpool, UK.
\bibitem[d]{haas}
Visitor from SLAC, Menlo Park, CA, USA.
\bibitem[e]{luna-garcia}
Visitor from Centro de Investigacion en Computacion - IPN,
  Mexico City, Mexico.
\bibitem[f]{podesta-lerma}
Visitor from ECFM, Universidad Autonoma de Sinaloa, Culiac\'an, Mexico.
\bibitem[g]{weber}
Visitor from Universit{\"a}t Bern, Bern, Switzerland.
\bibitem[h]{wenger}
Visitor from Universit{\"a}t Z{\"u}rich, Z{\"u}rich, Switzerland.

%
\vskip 0.25cm
  \bibitem{theorywidth} J.L. Rosner, M.P. Worah, and T. Takeuchi,
Phys. Rev. D {\bf 49}, 1363 (1994).
\bibitem{pdg} C. Amsler {\it et al.} (Particle Data Group), Phys. Lett. B {\bf 667}, 1 (2008).
\bibitem{wmasscombo}
  The Tevatron Electroweak Working Group (CDF and D0 Collaborations), arXiv:0908.1374 [hep-ex] (2009).
\bibitem{width_corr}
  V. Barger {\it et al.}, Phys. Rev. D {\bf 28}, 2912 (1983); M. Drees, C.S. Kim, and X. Tata, Phys. Rev. D {\bf 37}, 784 (1988).
\bibitem{CDFwidth}
  T. Affolder {\it et al.} (CDF Collaboration), Phys. Rev. Lett. {\bf 85}, 3347 (2000).
\bibitem{CDFwidth2}
 T. Aaltonen {\it et al.} (CDF Collaboration), Phys. Rev. Lett. {\bf 100}, 071801 (2008).
\bibitem{d0width}
  V.M. Abazov {\it et al.} (D0 Collaboration), Phys. Rev. D {\bf 66}, 032008 (2002).
\bibitem{combine} V.M. Abazov {\it et al.} (CDF and D0 Collaborations), Phys. Rev. D {\bf 70}, 092008 (2004).
\bibitem{LEPwidth} 
  S. Schael {\it et al.} (ALEPH Collaboration), Eur. Phys. J. C {\bf 47}, 309 (2006); G. Abbiendi {\it et al.} (OPAL Collaboration), Eur. Phys. J. C {\bf 45}, 307 (2006); P. Achard {\it et al.} (L3 Collaboration), Eur. Phys. J. C {\bf 45}, 569 (2006); P. Abreu {\it et al.} (DELPHI Collaboration), Eur. Phys. J. C {\bf 55}, 1 (2008).
\bibitem{d0det}
 V.M. Abazov {\it et al.} (D0 Collaboration), Nucl. Instrum. Methods A {\bf 565}, 463 (2006).
\bibitem{nim} V.M. Abazov {\it et al.} (D0 Collaboration), Nucl. Instrum. Methods A {\bf 609}, 250 (2009).
\bibitem{worldwmass} V.M. Abazov {\it et al.} (D0 Collaboration), Phys. Rev. Lett. {\bf 103}, 141801 (2009).
\bibitem{geo} The polar angle $\theta$ is defined with respect to
the positive $z$ axis, which is defined along the proton beam direction. Pseudorapidity 
is defined as $\eta=-\ln[\tan(\theta/2)]$.  
$\eta_{D}$ is the pseudorapidity measured with respect to the center of the detector.
\bibitem{bib:geant}
  R. Brun and F. Carminati, CERN Program Library Long Writeup W5013, 1993 (unpublished).
\bibitem{resbos} C. Balazs and C.P. Yuan, Phys. Rev. D {\bf 56}, 5558 (1997).
\bibitem{photos}
   E. Barberio and Z. Was, Comput. Phys. Commun {\bf 79}, 291 (1994); we use PHOTOS version 2.0.
\bibitem{cteq}
  J. Pumplin {\it et al.}, J. High Energy Phys. {\bf 07} (2002) 012.
\bibitem{lepzmass} R. Barate {\it et al.} (ALEPH Collaboration), Eur. Phys. J. C{\bf 14}, 1 (2000); P. Abreu {\it et al.} (DELPHI Collaboration), Eur. Phys. J. C{\bf 16}, 371 (2000); M. Acciarri {\it et al.} (L3 Collaboration), Eur. Phys. J. C{\bf 16}, 1 (2000); G. Abbiendi {\it et al.} (OPAL Collaboration), Eur. Phys. J. C{\bf 19}, 587 (2001); The ALEPH, DELPHI, L3, OPAL, SLD Collaborations, the LEP Electroweak Working Group, the SLD Electroweak and Heavy Flavor Groups, Phys. Rept. {\bf 427}, 257 (2006). 
\bibitem{ago} G. D'Agostini, Nucl. Instrum. Methods A {\bf 362}, 487 (1995).
\bibitem{matrix_method} V.M. Abazov {\sl et al.} (D0 Collaboration), Phys. Rev. D {\bf 74}, 112004 (2006). 
\bibitem{g2} F. Landry {\it et al.}, Phys. Rev. D {\bf 67}, 073016 (2003).
\bibitem{wgrad} U. Baur, S. Keller, and D. Wackeroth, Phys. Rev. D {\bf 59}, 013002 (1998).
\bibitem{zgrad}
  U. Baur, S. Keller, and W.K. Sakumoto, Phys. Rev. D {\bf 57}, 199 (1998); U. Baur {\it et al.}, Phys. Rev. D {\bf65}, 033007 (2002).

\end{thebibliography}
\end{document}